\documentclass[manuscript]{emulateapj}

\slugcomment{Accepted for publication in the Astrophysical Journal}

\shorttitle{Optical Spectroscopy of LBAS Blazars}
\shortauthors{M. S. Shaw et al}
\begin{document}

\title{Optical Spectroscopy of Bright Fermi LAT Blazars}

\author{Michael S. Shaw\altaffilmark{1}, Roger W.\ Romani\altaffilmark{1}, 
Stephen E.\ Healey\altaffilmark{1}, \\
Garret Cotter\altaffilmark{2}, Peter F.\ Michelson\altaffilmark{1}, \\
Anthony C.\ S.\ Readhead\altaffilmark{3}}

\altaffiltext{1}{Department of Physics/KIPAC, Stanford University, Stanford, CA 94305}
\altaffiltext{2}{Department of Astrophysics, University of Oxford, Oxford OX1 3RH, UK}
\altaffiltext{3}{Department of Astronomy, California Institute of Technology, Pasadena, CA 91125}

\begin{abstract}
	We report on HET and Palomar 5 m spectroscopy of recently identified
$\gamma$-ray blazars in
the {\it Fermi} LAT Bright Source List. These data provide identifications
for 10 newly discovered $\gamma$-ray flat spectrum radio quasars (FSRQ)
and six new BL Lacs plus improved spectroscopy for six additional BL Lacs. We  substantially improve the identification completeness of the bright
LAT blazars and give new redshifts and $z$ constraints, new estimates of the black hole
masses and new measurements of the optical SED.
\end{abstract}


\section{Introduction}

	The {\it Fermi} LAT pair production telescope has been surveying the
20\,MeV$-$300\,GeV $\gamma$-ray sky since 2008 August 11. Among the many
sources being detected by this mission, blazars dominate the extragalactic
sky. The first published set of these objects, the `LAT Bright 
AGN Sample' (LBAS) based on 3 months of sky survey exposure, included 132
bright ($\ge 10\sigma$) detections at $|b|>10^\circ$. The bulk (117) of these sources
have been associated with flat spectrum radio counterparts, which in turn
end up being mostly well-known flat spectrum radio quasars (FSRQ) and BL Lac
objects. A few additional sources are pulsars and a few remain unidentified.
The associations are based on the distribution of radio and X-ray properties of
the radio counterparts (CRATES; Healey et al. 2007) and the analysis allows 
a quantitative assessment of the probability of association.

	Even before the start of the {\it Fermi} mission, it was recognized
that a large sample of these blazars would be needed for source identification;
in the Candidate Gamma-Ray Blazar Survey (CGRaBS; Healey et al. 2008, and
references therein) we pursued optical spectroscopy of the sources selected to
be most similar to the previously known EGRET blazars. This set of 1625 blazars
included many sources without previous spectroscopic identification. Here we report
on previously unpublished optical spectra from our CRATES/CGRaBS survey and observations
for new sources from the LBAS list. The data allow classification of the blazar type
and redshift solutions for the strong-lined objects. Significant redshift constraints are
also obtained for the weak-lined BL Lacs. These measurements contribute ten FSRQ redshifts,
one new spectroscopic redshift for a BL Lac, and 11 additional BL Lac redshift constraints. 
Including these new data, 91\% of all LBAS sources have been spectroscopically typed, 84\%
have redshifts, and 70\% of the BL Lac have spectroscopic redshifts. We also 
use our spectra to examine the black hole masses and optical continua
of these blazar sources.

In this paper, we assume an approximate concordance cosmology---$\Omega_m=0.3$, $\Omega_\Lambda=0.7$, and $H_0=70 $km s$^{-1}$ Mpc$^{-1}$.

\section{Observations and Data Analysis}

\subsection{HET}

The bulk of the spectroscopic observations were performed with the 
9.2 m Hobby-Eberly Telescope (HET) at McDonald Observatory; many
of these observations were part of the CGRaBS survey \citep{cgrabs},
but additional data were taken on newly discovered LBAS sources.
The HET observes in the declination range $-11^\circ < \delta < +73^\circ$, and is 
fully queue scheduled \citep{she07}, allowing us to receive data remotely 
year round and to spread the cost of inclement weather and 
unfavorable conditions among the observing programs. We use the Marcario 
Low-Resolution Spectrograph, LRS 
\citep{hil98}, with grism G1 ($300$ lines mm$^{-1}$), a 2$^{\prime\prime}$ slit, and 
a Schott GG385 long-pass filter for a resolution of $R \approx 500$ 
between  $4150$ \AA\ and $10500$ \AA. Typical exposures are 
$2 \times 600 $ s for FSRQ objects and $2 \times 900 $ s for 
BL Lac objects, with the slit placed along the parallactic angle. 

\subsection{DBSP}

We also used the double spectrograph (DBSP) on the $5$\,m Hale Telescope 
at Palomar.  The observations with a 1$^{\prime\prime}$ slit at the parallactic angle
used a $600$ line mm$^{-1}$ grating on the blue side, covering
$\lambda\lambda3100-5200$  at a resolution of $\sim 2.8$ \AA\,
($R\sim 1450$). The red camera, with a $316$ line mm$^{-1}$ grating 
covers $\lambda\lambda5200-9500$, at a resolution $\sim 5.2$ \AA\,
($R\sim 1350$).  P200/DBSP is our primary tool for targets outside of the HET declination
range (i.e.  $-35^\circ<\delta<-11^\circ$ and $+73^\circ<\delta$)
and provides extra spectral coverage and resolution in the blue for bright BL Lacs.
Typical exposures were $2\times600$ s for FSRQs and up to 
$3\times900$ s for faint BL Lacs. 

\begin{figure*}
\epsscale{1.2}
\plotone{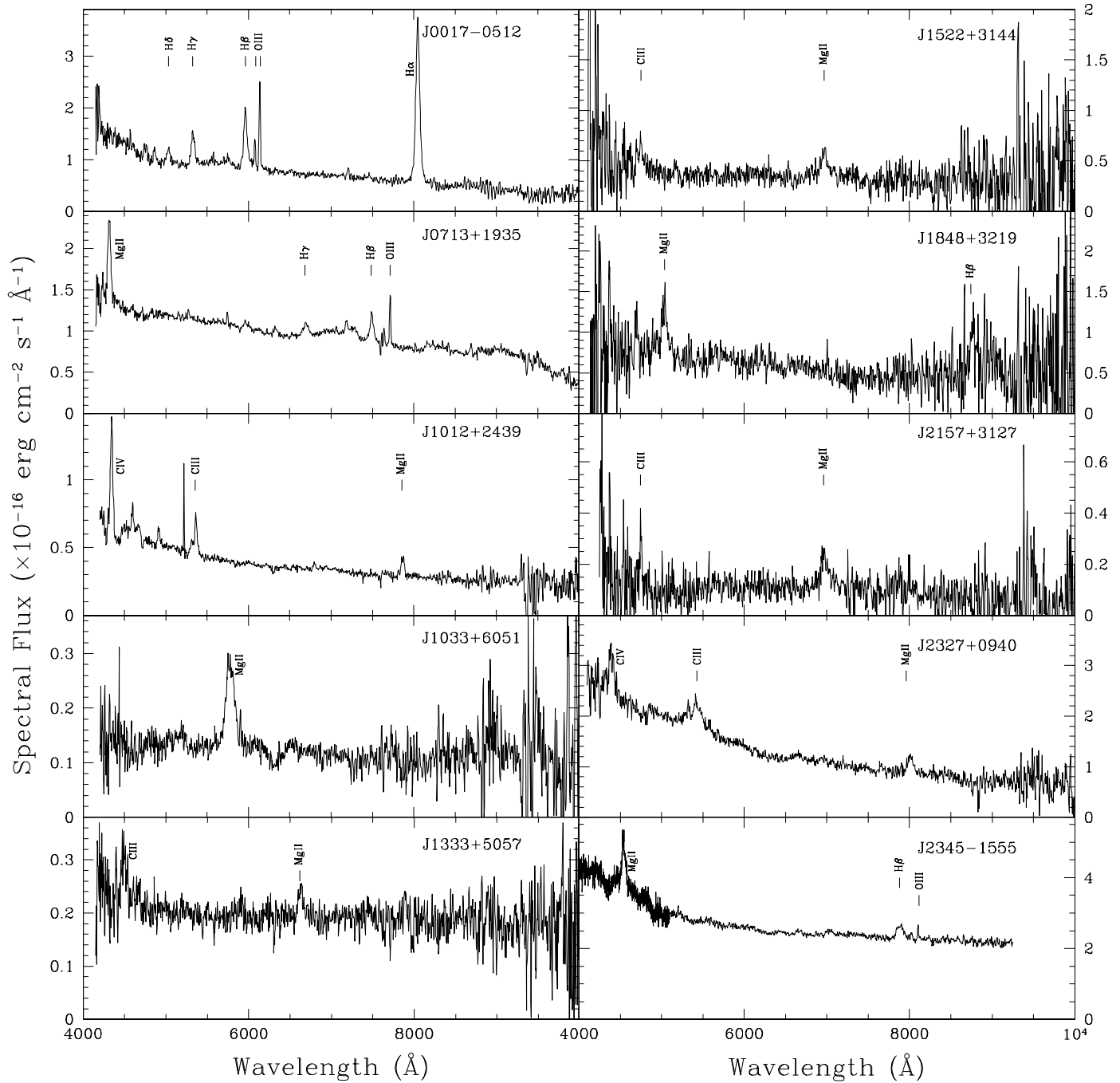}
\caption{New spectra of FSRQ objects in the {\it Fermi} LBAS sample.  The emission line properties are listed in Table 1.}
\end{figure*}

\subsection{Analysis Pipeline}

Data reduction was performed with the IRAF package \citep{tod86} using standard techniques. 
Wavelength calibration was performed with a neon-argon lamp at the HET 
and on the red side at DBSP, and iron-argon on the DBSP blue side. For
these relatively faint objects, we employ an optimal extraction 
algorithm \citep{val92} to maximize the S/N.

We perform spectrophotometric calibration using standard stars from \citet{oke90}.
In most cases the standard exposures from the data night were suitable,
but at the HET, standards from subsequent nights were sometimes used.
Due to differential slit losses and variable conditions between object and 
standard star exposures, we estimate that the accuracy of our absolute 
spectrophotometry is $\sim 30\%$ \citep{cgrabs}, although the relative spectrophotometry
is considerably better. Spectra are corrected for telluric absorptions
and visually cleaned of cosmic rays. Multiple exposures on a single target 
are combined into a single spectrum, weighting by S/N. 

We visually identify and measure emission line equivalent widths and FWHMs 
and derive redshifts from these line measurements. They are listed in Table 1. 
Line and redshift measurement techniques are discussed in \S 3.1.

\begin{deluxetable*}{lccccccccccccc}
\tabletypesize{\tiny}
\tablecaption{Line Properties for New LBAS FSRQ}
\tablehead{
\colhead{} & \colhead{} & \colhead{} & \colhead{} &\colhead{} & \colhead{} & \colhead{} &\multicolumn{2}{c}{H$\beta$} & \multicolumn{2}{c}{\ion{Mg}{2}} & \multicolumn{2}{c}{\ion{C}{4}} & \colhead{} \\
\colhead{Fermi Name} & \colhead{Name} & \colhead{RA} & \colhead{Dec} & \colhead{$F_{\nu,10^{14.7}}$} & \colhead{$\alpha$} &  \colhead{z} & \colhead{EW} &  \colhead{FWHM} & \colhead{EW} & \colhead{FWHM} & \colhead{EW} & \colhead{FWHM} & \colhead{Other Lines} \\
\colhead{} & \colhead{} & \colhead{} & \multicolumn{3}{c}{$10^{-28}$erg cm$^{-2}$s$^{-1}$Hz$^{-1}$}  & \colhead{} & \colhead{\AA$_\textrm{obs}$} & \colhead{kms$^{-1}_\textrm{rest}$}  & \colhead{\AA$_\textrm{obs}$} & \colhead{kms$^{-1}_\textrm{rest}$}  & \colhead{\AA$_\textrm{obs}$} & \colhead{kms$^{-1}_\textrm{rest}$} & \colhead{}
}
\startdata
OFGL J0017.4$-$0503 & J0017$-$0512 & 00 17 35.8 & $-$05 12 41.6 & 85 & 0.78$\pm$0.06 & 0.227 & 59 & 3009 & & & & & H$\alpha$,H$\gamma$,H$\delta$,\ion{O}{2}, \ion{O}{3}\\
OFGL J0714.2+1934 & J0713+1935 & 07 13 55.6 & 19 35 00.3 &  13 & 1.3$\pm$0.07 & 0.540 & 19 & 2815 & 30 & 2841 & & & H$\alpha$,H$\gamma$,H$\delta$,\ion{O}{2}, \ion{O}{3}\\
OFGL J1012.9+2435 & J1012+2439 & 10 12 41.3 & 24 39 23.3 & 4.6 & 1.0$\pm$0.05 & 1.805 &  &  & 24 & 2489 & 45 & 2360 & \ion{C}{3},1650,1730\\
OFGL J1034.0+6051 & J1032+6051 & 10 32 53.9 & 60 51 27.5 & 1.5 & 1.3$\pm$0.12  & 1.064 & &  & 140 & 6495 & & & \\
OFGL J1333.3+5088 & J1333+5057 & 13 33 53.7 & 50 57 35.9 & 2.3 & 1.9$\pm$0.09 & 1.362 & &  & 28 & 3147 & & & \ion{C}{3}\\
OFGL J1522.2+3143 & J1522+3144 & 15 22 09.9 & 31 44 14.3 & 4.3 & 1.8$\pm$0.18 & 1.487 & &  & 40 & 4086 & & & \ion{C}{3}\\
OFGL J1847.8+3223 & J1848+3219 & 18 48 22 & 32 19 02.6 & 7.5 & 0.71$\pm$0.16 & 0.798 & 75 & 2878 & 52 & 2682 & & & \ion{C}{2}\\
OFGL J2157.5+3125 & J2157+3127 & 21 57 28.8 & 31 27 01.3 & 1.4 & 1.5$\pm$0.60 & 1.486 & &  & 103 & 4185 & & & \ion{C}{3}\\
OFGL J2327.3+0947 & J2327+0940 & 23 27 33.5 & 09 40 09.5 & 16 & 0.51$\pm$0.07 & 1.843 & &  & 32 & 4840 & 24 & 6580 & \ion{C}{2}, \ion{C}{3}\\
OFGL J2345.5$-$1559 & J2345$-$1555 & 23 45 12.4 & $-$15 55 07.8 & 31 & 1.3$\pm$0.02 & 0.621 & 17 & 3835 & 17 & 3630 & & & \ion{O}{3}, H$\gamma$,H$\delta$\\
\enddata
\end{deluxetable*}
\bigskip

\section{Results}

We present spectra for ten FSRQ objects, six new BL Lac identifications and 
spectra, and high S/N spectroscopy on six additional objects previously known
to be BL Lacs. We present spectroscopic redshifts for all ten FSRQ, and for 
one BL Lac.  We derive virial BH masses for the FSRQ, and for BL Lacs that lack 
spectroscopic redshifts, we extract significant constraints on the redshift. 

\subsection{FSRQ Spectra}

We present ten previously unpublished FSRQ spectra with redshifts ranging 
from 0.227 to 1.805. These redshifts were previously listed in \citet{lbas}. 
All redshifts are confirmed by multiple emission lines and derived by 
cross-correlation analysis using the rvsao package \citep{rvsao}.
Table 1 gives the approximate continuum fluxes 
at $10^{14.7}$Hz ($\lambda= 5981$ \AA) and spectral indices $\alpha$
($F_\nu \propto \nu^{-\alpha}$). The fluxes, as above, have uncertainties 
as large as 30\% due to unknown slit losses. The spectral index uncertainties are
estimated by fits to independent subsets of the spectral range. 

The properties of strong emission lines are listed in Table 1. We are 
particularly interested in the broad emission lines H$\beta$, \ion{Mg}{2}, and \ion{C}{4}, 
as their virialized broad components are used to make estimates of the
black hole mass. 
For these lines, we take care to measure the Gaussian full width at half maximum (FWHM) of the 
broad component, avoiding contamination by other spectral features. 

For H$\beta$, it is important to include narrow components
in the line fit. We fit the continuum to a power law, and, following 
\citet{md04}, we simultaneously fit broad and narrow H$\beta$, 
narrow [\ion{O}{3}]4959 \AA, and narrow [\ion{O}{3}]5007 \AA. We require rest 
wavelengths of narrow H$\beta$ and the [\ion{O}{3}] lines to match 
laboratory values; the broad H$\beta$ wavelength was free. All lines are modeled 
with Gaussian profiles. The continuum is measured at 5100 \AA.

For \ion{Mg}{2}, we first subtract off our power law fit. Then, we subtract 
a template \citep{tsu06} of the broad \ion{Fe}{2} complexes near 2800 \AA, 
to minimize contamination with \ion{Mg}{2}. We then fit the \ion{Mg}{2} line 
itself with a broad+narrow Gaussian. For this line the continuum 
measurement is, as usual, made at 3000 \AA, to avoid Fe contamination \citep{md04}.

\ion{C}{4} lines sometimes suffer strong associated absorption, but such corrections were
modest for the few cases presented here. Also, for consistency with the SDSS
measurements (see below) and to minimize sensitivity to associated absorbers we 
report here the FHWM of the actual \ion{C}{4} line \citep{she08}.

\subsection{Black Hole Mass Estimates}

We can use the emission line measurements reported above to estimate black hole 
masses for the FSRQ objects in our sample using the empirical 
`virial' scaling relationships \citep{mj02}. Mass scaling relations have been 
applied to large samples of quasars in the past and are calibrated by 
reverberation mapping techniques \citep{ves09}.

\begin{figure}[ht!]
 \epsscale{1.2}
 \plotone{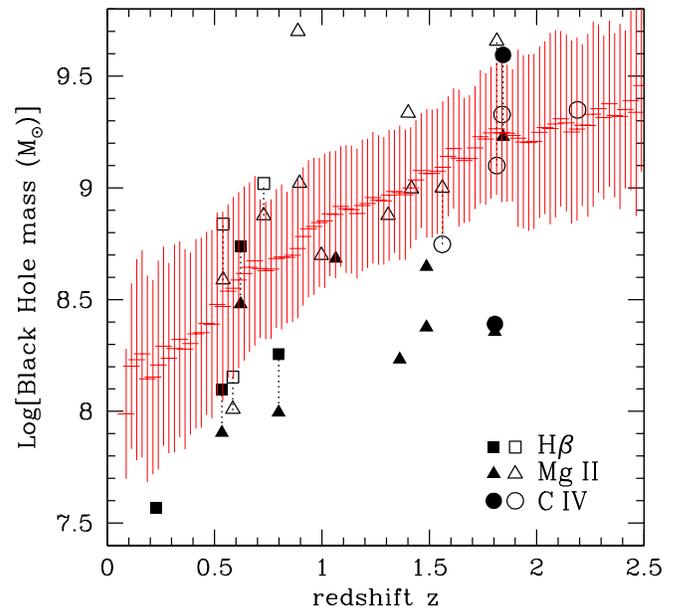}
 \caption{BH mass estimates as a function of redshift for LBAS sources. 
Filled points are measured from FSRQ spectra presented in this paper, open points
are for spectra of LBAS FSRQ available from the SDSS archive
(Squares: $H\beta$ $\lambda$4861, Triangles: \ion{Mg}{2} $\lambda$2800, Circles: 
\ion{C}{4} $\lambda$1550). When two species are used for a given source,
dashed lines connect the mass estimates.
For comparison, the mean and rms range of mass estimates
from the full SDSS catalog \citep{she08} are shown.  \label{fig:bh}} 
\end{figure}

Continuum flux measurements were made for each line at the appropriate
rest wavelength ($5100$ \AA\ for H$\beta$, $3000$ \AA\ for \ion{Mg}{2}, $1350$ \AA\ for \ion{C}{4}), and converted to continuum luminosity $\lambda L_\lambda$, using our assumed cosmology. 
For two objects 
in our sample, J1012+2439 and J2327+0940, the \ion{C}{4} line is clearly present, but the 
spectra do not extend blue-ward enough to cover $1350$ \AA; for these, we extrapolate
a power law fit to the continuum at longer wavelengths. We follow 
\citet{she08} in the calculation of the black hole masses:
\begin{equation}
 \log\left(\frac{M}{M_\odot}\right) = 
a + b\cdot \log\left(\frac{\lambda L_\lambda}{10^{44}\textrm{erg s}^{-1}}\right) + 
2 \log\left(\frac{\textrm{FWHM}}{\textrm{km s}^{-1}}\right)
\end{equation}
where $a = 0.505$, $b=0.62$ for \ion{Mg}{2}; $a=0.672$, $b=0.61$ for H$\beta$; and 
$a=0.66$, $b=0.53$ for \ion{C}{4}.

\begin{figure*}
\epsscale{1.2}
\plotone{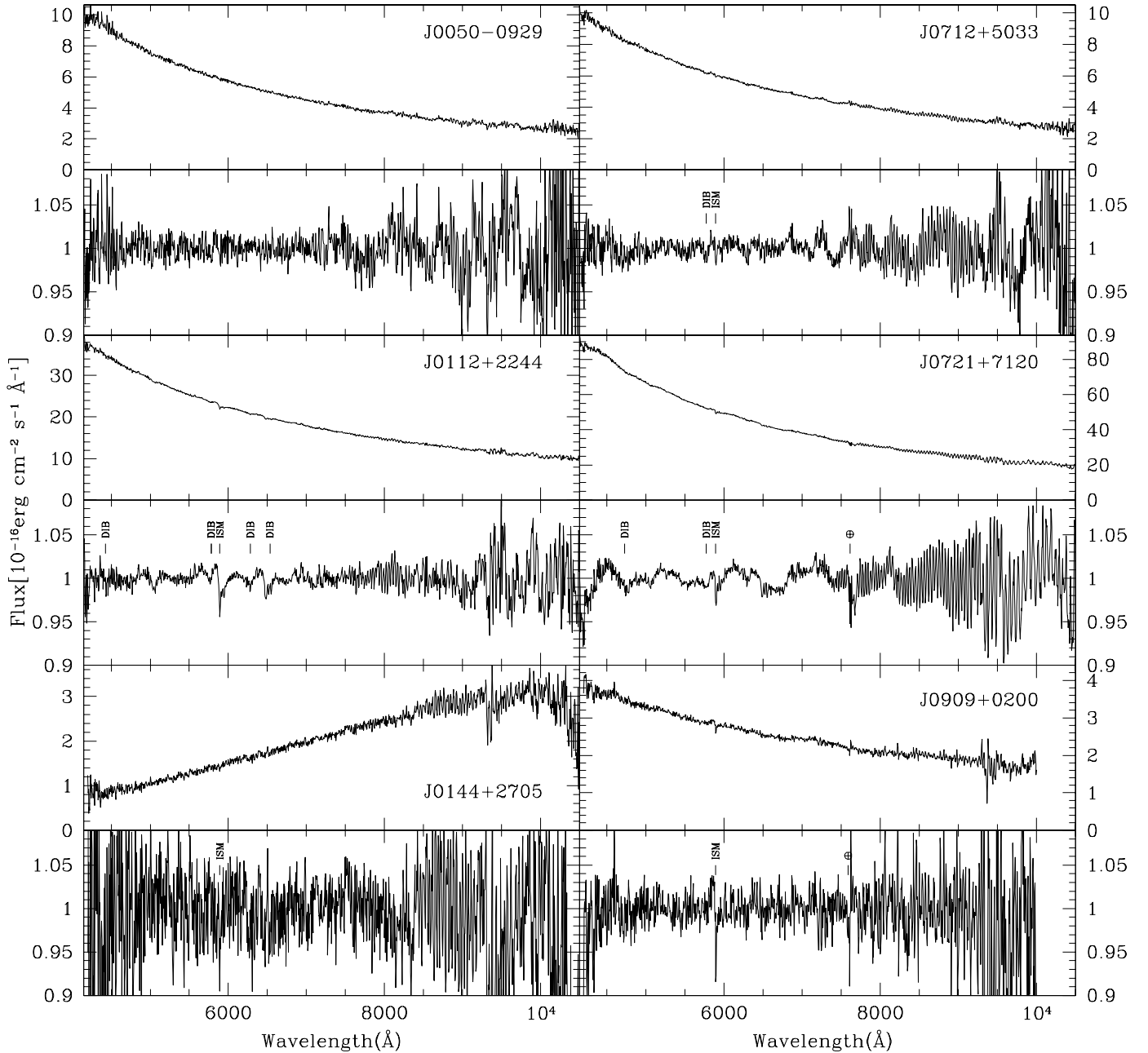}
\caption{New spectra of LBAS BL Lac objects. Top panels are flux-calibrated spectra 
with flux in $10^{-16}$erg cm$^{-2}$ s$^{-1}$  \AA$^{-1}$. 
Bottom panels show spectra normalized by dividing out the continua.  Telluric and 
interstellar absorption features are labeled.}
\label{fig:bll1}
\end{figure*}

Masses are plotted in Figure \ref{fig:bh}. Values measured from spectra presented 
in this paper are indicated by filled points (species used indicated by point style). 
Some of the LBAS FSRQ have spectra in the SDSS archive \citep{sch07}.  We recovered these and measured  line profiles and continuum fluxes exactly as for our new objects to obtain similar 
BH mass estimates (open points). For about half of these sources, SDSS published independently estimated values for $M$ \citep{she08}. We find adequate agreement, with rms differences in $\log(M)$ of
0.04 (H$\beta$) and 0.14 (\ion{Mg}{2}, \ion{C}{4}), comparable to the inter-species dispersion.

In general, the virial mass estimators are
of limited accuracy for individual objects. As one illustration, when estimates
are available for a given object from more than one species, we connect the points with
a dashed line.  The pipeline processing of the entire Sloan DR5 sample of optically 
selected quasars provides a large number of such mass estimates. In Figure \ref{fig:bh}
we show the mean and rms $\log(M)$ values as a function of redshift. For both the
$\gamma$-ray selected sample presented here and the optical SDSS sample, the mean mass
increases with redshift, likely a simple flux bias. However, we note that the 
$\gamma$-ray sources trend about $3\times$ lower in mass than the optically 
selected objects, albeit with large scatter.
There are several possible origins of this trend. One possibility is that
this is a result of the fainter optical limit in the follow-up
to the LBAS sample.  Another selection effect may be imposed by the $\gamma$-ray 
selection's dependence on the jet flux, which, as a strong function of the 
viewing angle to the relativistic jet, is more weakly dependent on the underlying source 
luminosity and presumably black hole mass. 
A final possibility stems from orientation: virial estimates (Equation 1) assume
isotropic velocities in the broad line region. However, blazars are known to have
the jet axis, and hence the black hole spin and inner accretion disk axis, close to
the Earth line-of-sight. If the broad line region has a toroidal structure \citep{fi08}, then
blazars will have lines with FWHM decreased by $\langle {\rm sin}\,i \rangle$ and lower virial
mass estimates than QSOs viewed at larger inclination angles.  Larger data sets from 
future LAT samples should allow us to probe the reality and origin of such effects.

\begin{figure*}
\epsscale{1.2}
\plotone{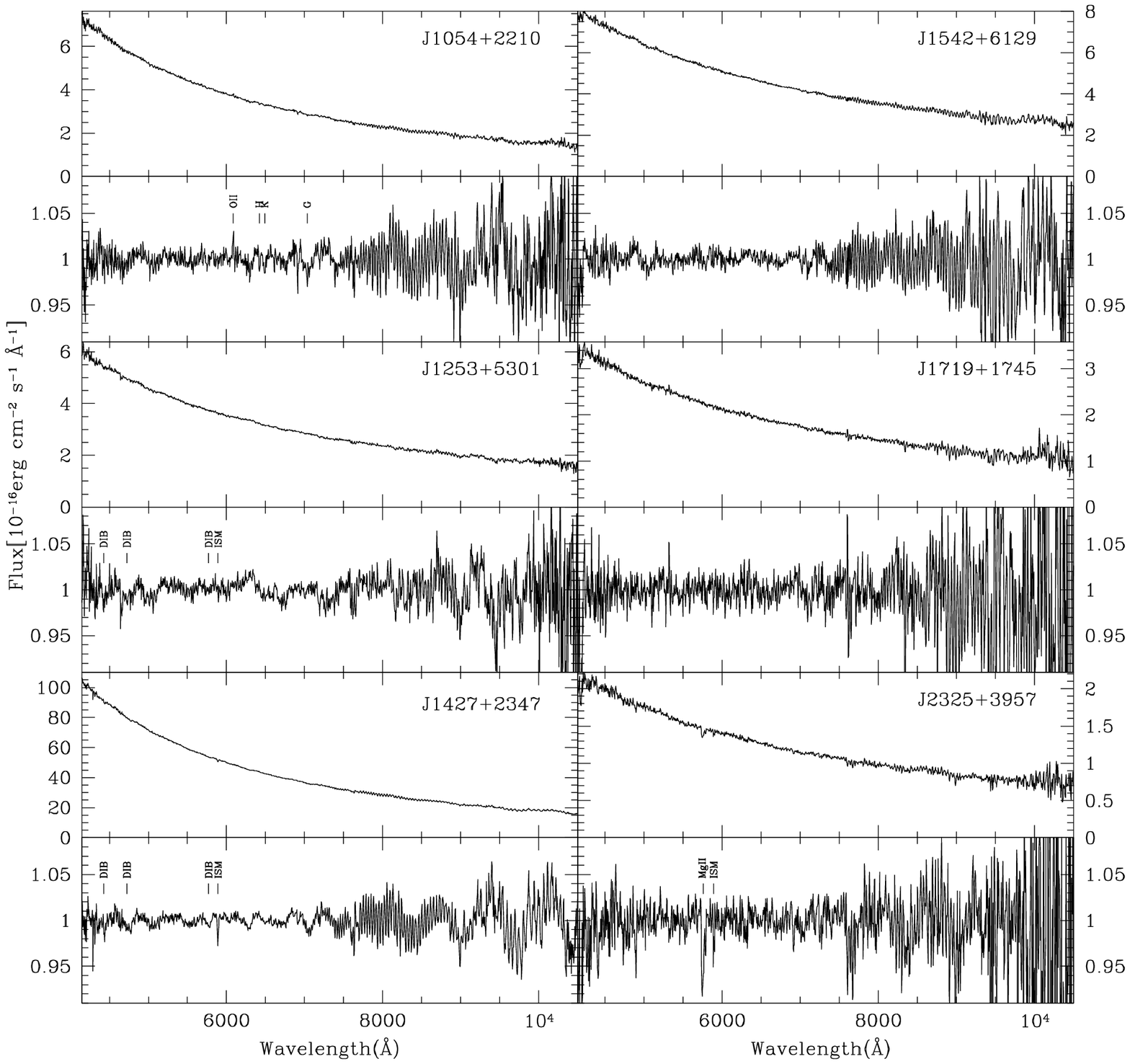}
\caption{As for Figure 3. J1054+2210 has a redshift solution $z=$0.634. J2325+3957
has an intervening absorption line system at $z=$1.04.} 
\label{fig:bll2}
\end{figure*}

\subsection{BL Lac Properties}

We have obtained high sensitivity (S/N $\sim$ 100) spectra of twelve LBAS BL Lacs. 
Following \citet{mar96} and \citet{cgrabs}, we adopt here the pragmatic
`optical spectroscopic' definition of a BL Lac as a blazar lacking emission lines 
with observed equivalent width greater than $5$ \AA\ and a limit on any possible 
4000 \AA\ spectral break of 
$<40$\%. For some sources in CGRaBS we have seen that, as the continuum 
level fluctuates, the source passes from spectroscopic BL Lac to spectroscopic FSRQ.
We designate a source as a BL Lac if during any of our 
observations with adequate spectral range it has satisfied the spectroscopic
definition. In this connection, one of the high-mass objects appearing in
Figure \ref{fig:bh} should be mentioned. This is J1058+0133, designated an FSRQ 
in LBAS based on a clear detection of a broad \ion{Mg}{2} line at z=0.888. However, 
this line has an EW of only 3.5 \AA; accordingly this blazar should be formally 
designated a BL Lac 
at the epoch of the SDSS spectrum. The very strong non-thermal continuum, in fact, 
drives up the apparent luminosity and the BH mass estimate. This emphasizes that 
non-thermal continuum can artificially increase the apparent mass; the virial 
estimates are not calibrated for BL Lac sources.

	The spectral range requirement for BL Lac designation is that the spectrum covers
$[\lambda_\textrm{min}$ to $\lambda_\textrm{max}]$ with sufficient S/N to detect
at least one of the standard AGN broad emission lines --- Ly$\alpha$, \ion{C}{4}, \ion{C}{3}, 
\ion{Mg}{2}, H$\beta$, or H$\alpha$.  In practice this means 
adequate S/N over ${\lambda_\textrm{max}}/{\lambda_\textrm{min}} > 1.75$;
this is easily satisfied for the BL Lacs in our LBAS sample.

To describe the BL Lac continua, we fit the spectra to a power law, 
using the IRAF nfit1d routine. Most are well fit with rising (blue)
power-laws, suggesting a synchrotron $\nu F_\nu$ peak well above the
optical band. J0144+2705 is, however, very red. This is confirmed by 
USNO B1/2MASS fluxes which rise to the J-band and flatten above 1.5$\mu$m.
The Galactic extinction in the direction of this source is $A_B = 0.3$\,mag, so 
local absorption is unlikely to redden the source. It seems plausible that
this BL Lac has a low synchrotron peak frequency $\nu_\textrm{max} \sim 10^{14}$Hz.
This may be tested by broader study of the SED.

The spectral indices ($F_\nu \propto \nu^{-\alpha}$) and fluxes of the BL Lac
are reported in Table \ref{tab:bll}.  For the continuum normalization we 
assume the $30$\% absolute spectrophotometric uncertainty noted above; for 
the spectral indices we again estimate the dominant systematic errors from 
fits to non-overlapping 
sections of the continuum. The continuum-normalized spectra are shown in
Figures 3 \& 4. These plots also label notable interstellar features e.g. 
Na $\lambda 5892$, and the diffuse interstellar bands, or DIBs \citep{dib}.

\subsection{BL Lac Redshift Limits}

The strong dominance of the non-thermal nuclear continuum in BL Lacs guarantees that
the equivalent widths of any broad nuclear lines or absorption features from the
host galaxy are very small. While recent work with 8 m-class telescopes has provided
some redshift solutions, spectroscopic redshifts for many objects remain elusive
\citep{sba1}. Nevertheless,
quantitative limits on host spectral features can be useful in extracting redshift
constraints. For example, based on the lack of a Ly$\alpha$ forest-induced break 
in the HET spectra, none of our BL Lacs can have $z>2.41$. This rather weak bound can
be improved with UV coverage beyond the HET LRS 4150 \AA\, limit.

	Also, there is good evidence, especially from HST imaging, that BL Lacs
reside in relatively uniform giant elliptical hosts, with  $M \sim -22.9$ \citep{sba1}.
Analysis to limit the contribution of such a host to the observed spectrum can
thus provide a lower limit on the BL Lac redshift \citep{sba2}. We develop
here a similar spectral bound on redshift, using the Ca H/K and G-band absorption
features, the strongest narrow features in the host spectrum.

\begin{deluxetable*}{lcccccc}
\tablecolumns{7}
\tabletypesize{\small}
\tablecaption{BL Lac Continuum properties for LBAS sources\label{tab:bll}}
\tablehead{
\colhead{Fermi Name} & \colhead{Name} &  \colhead{RA} & \colhead{dec} & \colhead{$F_{\nu,10^{14.7}}$} &  \colhead{$\alpha$} & \colhead{z} \\
\colhead{} & \colhead{} & \colhead{} & \multicolumn{3}{c}{$10^{-28}$ergcm$^{-2}$s$^{-1}$Hz$^{-1}$} & \colhead{}
}
\startdata
OFGL J0050.5-0928 & J0050$-$0929 & 0 50 41.3 & $-$9 29 5.1 & 69 & 0.48$\pm$0.01 & $> 0.44$ \\
OFGL J0112.1+2247 & J0112+2244 & 01 12 05.8 & 22 44 38.8 & 270 & 0.54$\pm$0.01 & $> 0.24$ \\
OFGL J0144.5+2709 & J0144+2705 & 01 44 33.5 & 27 05 03.0 & 17 & 3.6$\pm$0.06 & $> 0.45$ \\
OFGL J0712.9+5034 & J0712+5033 & 07 12 43.6 & 50 33 22.6 & 71 & 0.56$\pm$0.02 & $> 0.47$ \\
OFGL J0722.0+7120 & J0721+7120 & 07 21 53.4 & 71 20 36.4 & 590 & 0.31$\pm$0.02 & $> 0.06$ \\
OFGL J0909.7+0145 & J0909+0200 & 09 09 39.85 & 02 00 05.3 & 33 & 1.1$\pm$0.03 & $> 0.54$ \\
OFGL J1054.5+2212 & J1054+2210 & 10 54 30.6 & 22 10 54.8 & 46 & 0.24$\pm$0.02 & $> 0.60^\dagger$ \\
OFGL J1253.4+5300 & J1253+5301 & 12 53 11.9 & 53 01 11.6 & 42 & 0.56$\pm$0.01 & $> 0.77$ \\
OFGL J1427.1+2347 & J1427+2347 & 14 27 00.3 & 23 47 59.9 & 600 & 0.01$\pm$0.01 & $> 0.03$\\
OFGL J1543.1+6130 & J1542+6129 & 15 42 56.9 & 61 29 55.3 & 61 & 0.74$\pm$0.02 & $> 0.63$ \\
OFGL J1719.3+1746 & J1719+1745 & 17 19 13 & 17 45 06.3 & 25 & 0.67$\pm$0.03 & $> 0.58$ \\
OFGL J2325.3+3959 & J2325+3957 & 23 25 17.8 & 39 57 36.5 & 17 & 0.76$\pm$0.02 & $> 1.05$ \\
\enddata
\tablecomments{$\dagger$ Measured redshift $z=0.634$.}
\end{deluxetable*}
\bigskip

	We first determine the expected equivalent width of a given absorption line
by measuring an elliptical template spectrum and then computing the
net equivalent width in the observed host+nucleus spectrum at each redshift 
following \citet{sba2}:
\begin{equation}
EW_\textrm{expected} = \frac{(1+z)EW_\textrm{h}}{1 + F_\textrm{n} / F_\textrm{h}}
\end{equation}
where $EW_\textrm{h}$ is the equivalent width of the line in the host galaxy spectrum, $F_\textrm{n}$ is the 
nuclear flux at the wavelength of that absorption line, and $F_\textrm{h}$ is the flux of the 
giant elliptical host at that wavelength. $F_\textrm{h}$ is corrected for the slit/extraction
aperture losses, assuming a deVaucoleurs profile with redshift-dependent angular size 
set by a constant radius $r_0=10$\,kpc. $F_\textrm{n}$ is the observed flux minus that expected 
from the host in our extraction aperture. This predicted feature strength is computed
for the major absorption lines at each redshift.

	We next compute the 3$\sigma$ limit on narrow absorption features in
the observed spectrum as a function of wavelength from the local rms of negative
spectral fluctuations, after high-pass filtering to remove broad continuum features
and after exclusion of zero-redshift Galactic and telluric features. We find that
the line sensitivity varies considerably across the spectrum (Sbarufatti et al. 2006
assume constant sensitivity for their VLT spectra). For the HET data, limited 
blue sensitivity weakens the
constraints below $\sim 5000$ \AA, while strong fringing (uncorrectable due the varying
HET pupil) weakens the bounds beyond $\sim 8000$ \AA. The sensitivity also varies 
due to the changing resolution; this is especially important for the DBSP data, although
no DBSP spectra are used in the measurements of the BL Lacs presented here. 

\begin{figure}
\epsscale{1.1}
\plotone{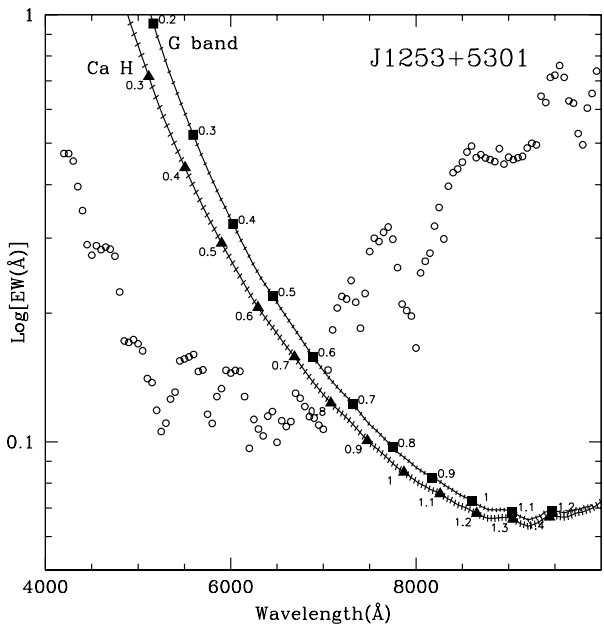}
\caption{Expected equivalent widths of the H (3934 \AA, triangles) and G 
(4304 \AA, squares) features
calculated at different redshifts. The 3$\sigma$ limit on line equivalent widths 
across the spectrum is plotted with open circles. As long as either H or G is 
above the circles, the corresponding redshift is excluded. Here, we excluded $z<0.77$.} 
\label{fig:minz}
\end{figure}

	Comparison of the expected and limiting EW curves gives a lower limit
on the BL Lac redshift (for the assumed standard host magnitude) whenever one
absorption feature is excluded at the 3$\sigma$ level. Generally the Ca features
provide the strongest constraint. However, since our HET spectra are limited to $\lambda
>4150$ \AA, the G-band at $\lambda=4304$ \AA\, is used to exclude $z\le 0.05$.
Figure \ref{fig:minz} shows the exclusion curve for J1253+5301, where we find
$z> 0.77$. The results for the BL Lac data are presented in Table \ref{tab:bll}.
With our HET data, we can exclude redshifts less than $z \sim 0.5 - 0.8$.
For a few particularly bright BL Lac, however, the redshift constraints are substantially weaker as the expected equivalent widths are smaller.
Higher resolution, higher S/N spectra can, of course, improve these bounds.

	A few objects deserve particular comment. For J1054+2210, after completing the
redshift limit
analysis, weak H/K and G-band features plus \ion{O}{2} $\lambda 3727$ emission were
found at redshift $z=0.634$. This detection is confirmed by cross-correlation 
analysis. The measured redshift
is just above the redshift bound of $z>0.60$ given by the exclusion analysis. For J2325+3957,
the EW limit analysis gave a redshift bound of $z>0.77$. We have detected
an intergalactic absorption system in this spectrum, with \ion{Mg}{2} doublet and \ion{Fe}{2} absorption features corresponding to $z=1.04$. This stronger bound, listed
in Table \ref{tab:bll}, is consistent with the EW limit analysis.

\section{Conclusions}

	The optical spectroscopy presented here provides significant new information
on some of the less well-known members of the set of the LBAS $\gamma$-ray blazars.

	For all of the new FSRQ, precision redshifts are measured from the
spectra. For one of the new BL Lacs, we determine the redshift spectroscopically;
for the others we extract limits on the redshift from the spectrum. Continuum
fluxes and spectral indices are provided that will be useful for SED plots.
These measurements bring the LBAS to 84\% redshift completion.
For most of the unsolved BL Lacs we extract significant lower limits on the redshift.
These will be helpful in modeling the $\gamma$-ray BL Lac population and evolution
and in focusing redshift searches to a narrower spectral range.

For the strong-line objects we present estimates of the virial black hole masses.
Although such mass estimates have large uncertainties, our objects appear
to have somewhat lower mass than typical for the SDSS quasar sample. Like 
that sample, the mean mass increases with redshift. In the one case where
a BL Lac-like object has a mass measurement, the excess continuum artificially inflates
the apparent black hole mass.

	The spectroscopic analysis presented here for the small LBAS blazar sample
will be much more powerful when applied to the full sample of LAT blazars
detected during the first year sky survey, which should include 500-1000 objects.
We continue to collect spectroscopic data on the LAT blazars as they are detected
and plan to use the optical analysis of this larger data set to probe the
evolution of the $\gamma$-ray emitting objects.

\acknowledgments

The Hobby*Eberly Telescope (HET) is a joint project of the University of Texas at
Austin, the Pennsylvania State University, Stanford University, Ludwig-Maximilians-Universitaet
Muenchen, and Georg-August-Universitaet Goettingen. The HET is named in honor of its principal
benefactors, William P. Hobby and Robert E. Eberly.
The Marcario Low Resolution Spectrograph is named for Mike Marcario of High Lonesome
Optics, who fabricated several optics for the instrument but died before its completion.
The LRS is a joint project of the Hobby*Eberly Telescope partnership and the Instituto de
Astronomıa de la Universidad Nacional Autonoma de Mexico.

MSS and SEH are supported by NASA under grants NNX08AW30G and NAS5-00147. ACSR is supported by the NSF under grant AST-0808050.

\end{document}